Reza Ghaiumy Anaraky*, Byron Lowens, Yao Li, Kaileigh A. Byrne, Marten Risius, Xinru Page, Pamela Wisniewski, Masoumeh Soleimani, Morteza Soltani, and Bart Knijnenburg

# Older and younger adults are influenced differently by dark pattern designs

Abstract:
Online websites often use dark patterns to increase users' information disclosure. Common examples include "opt-out" privacy defaults, positive framing, and positive justification messages encouraging disclosure behavior. Considering that prior research has found older users undergo a different privacy decision-making process compared to younger adults, more research is needed to inform the behavioral privacy disclosure effects of these strategies for different age groups. To address this gap, we used an existing dataset of an experiment with a photo-tagging Facebook application. This experiment had a 2x2x5 between-subjects design where the manipulations were common dark pattern design strategies: framing (positive vs. negative), privacy defaults (opt-in vs. opt-out), and justification messages (positive normative, negative normative, positive rationale, negative rationale, none). We compared older (above 65 years old, N=44) and young adults (18 to 25 years old, N=162) privacy concerns and disclosure behaviors (i.e., accepting or refusing automated photo tagging) in the scope of dark pattern design. Overall, we find support for the effectiveness of dark pattern designs in the sense that positive framing and opt-out privacy defaults significantly increased disclosure behavior, while negative justification messages significantly decreased privacy concerns. Regarding older adults, our results show that certain dark patterns do lead to more disclosure than for younger adults, but also to increased privacy concerns for older adults than for younger. However, there was no influence of these concerns on disclosure, and instead, they are outweighed by the prodisclosure effects of dark patterns. This suggests that privacy concerns may not be a sufficient force to drive individuals to act on protecting their privacy when in the presence of dark patterns and that such patterns may be even more dangerous for older users. We discuss the implications of this work.

## 1 Introduction

The Human-Computer Interaction (HCI) and networked privacy literature predominantly present a deficit-based narrative for older adults' technology use and privacy outcomes. This narrative frames older adults as individuals who have difficulty keeping up with the technology [82], managing their digital privacy [14, 64], and protecting themselves against privacy threats [86] compared to young adults. However, recent developments in the fields of HCI and privacy counter this deficit-based narrative by pointing out that older adults may have a different decision-making process rather than a sub-optimal one, and technologies should be designed with that in mind [3, 23]. (This approach includes considering the potential benefits of biometric authentication methods, which can offer older adults a more user-friendly and secure way to access digital services [50].)For example, Knowles and Hanson [42] argue that older adults' non-use of technology results from their high level of privacy concerns and, therefore, non-use is a well-informed decision. Moreover, Anaraky et al. [3] compared young and older adults' privacy disclosure decision-making processes and found that older adults make more calculated privacy choices than younger adults.

Privacy disclosure decisions, however, are not always dependent on the characteristics of the end-user and can be influenced by external manipulations. System designers may use design heuristics to nudge users to disclose their data [4, 30, 32, 43, 58]. In marketing and e-commerce research, maximizing disclosure is a goal that has led to the creation of "dark pattern" designs, where nudges are used as design interventions to increase disclosure [12]. For example, designers may configure a setting to disclose-by-default (opt-out) and present a biased positive framing when presenting choice options [25, 53, 54], or add persuasive messages to motivate users to disclose more information [40].

*Corresponding Author:
**Reza Ghaiumy Anaraky:** New York University,
E-mail: rg4598@nyu.edu

Given that older adults have been shown to have a very different privacy decision-making process compared to young adults, we used an existing dataset from Ghaiumy Anaraky et al. [4] to investigate the effect of dark pattern design on older and younger adults' privacy attitudes and behaviors. Ghaiumy Anaraky et al.

[4] initially used this dataset to examine the impact of justification messages (as norms) on the compliance-inducing effects of framing and defaults. They found that the presence of any justification messages increases the default effect.(Moreover, ensuring that personal and confidential data is appropriately classified not only enhances data management but also safeguards individuals' privacy rights [75].) In the present work, we conceptualize framing, defaults, and justification messages as dark pattern design strategies and situate this work in the dark pattern design literature. We further contribute to the privacy literature by investigating how these design patterns influence older and young adults differently. Therefore, we pursue the following research questions:

*RQ1*: What are the effects of dark pattern designs on users' attitudes and behaviors?

*RQ2*: How do older adults react differently than young adults to dark pattern designs in terms of their attitudes and behaviors?

To study individuals' privacy decision mechanisms in the context of dark pattern designs, we used an existing dataset of an experiment in which participants could purportedly automatically tag themselves in their friends' photos, and their friends could be tagged in the user's own photos by using a Facebook application. Previous literature studies tagging decisions as privacy decisions [9, 49, 85, 95]. Tagged photos appear on the timeline of the tagged person, thereby explicitly identifying the person in the photo and sharing the photo with that person's contacts. The study involved a 2x2x5 betweensubjects experimental design where the manipulations were three common means by which dark-pattern designs encourage disclosures [12]: 1) the framing of the decision (i.e., "tag me in the photos" vs. "do not tag me in the photos"), 2) the default setting (i.e., opt-out vs. opt-in), and 3) the use of a justification message (a message that could either encourage or discourage users from using the application). After their interaction with the manipulated dark-pattern designs, participants indicated their privacy concern [40, 48] and tagging decision (accept to tag vs. refuse to tag) as the attitudinal and behavioral dependant variables. To enable a comparison between older and younger participants, we followed established age criteria in the literature: we considered individuals aged 65 and above as older adults [15] and limited our young adult sample to college-age adults, between 18-25 years of age [60, 62].

**Contributions:** This study was unique because it was carried out in an ecologically valid setting on the user's own Facebook account. Our results show that older adults had heightened privacy concerns when faced with an opt-out default than young adults. However, when it came to actual behavior, despite this privacy concern, an increase in disclosure was still observed. Furthermore, we found that older adults are more likely to be influenced by framing nudges than young adults. In light of these findings, we provide design suggestions for supporting older adults' privacy.

## 2 Background

In this section we first summarize the literature on older and younger adults' privacy decisions. We then discuss dark pattern designs and common strategies to maximize users' disclosure (namely framing, defaults, and justification messages).

### 2.1 Privacy Decisions of Older Adults vs. Younger Adults

A large body of literature investigating age-related differences in digital privacy identifies older adults as individuals who experience more difficulties than younger adults in managing their digital privacy (e.g., [11, 14, 42, 64, 90]. Some argue that older adults are less likely to protect themselves against privacy-related risks [86, 98]. Lack of awareness of the privacy risks has been cited as a critical factor impacting older adults' privacy decisions [46]. For example, age-related differences have been found in research investigating content sharing and sociability and how these components are associated with the need for privacy among Facebook users [14]. Researchers discovered that younger adults are more competent in their Facebook usage and are more informed about and able to make changes to their privacy settings. In contrast, older adults seemed to have difficulties understanding the privacy settings and be less aware of social privacy issues.

Psychological literature corroborates how older and younger adults exhibit fundamental behavioral distinc-tions in their decision-making patterns, encompassing differences in risk preference and reliance on goal-driven approaches [96]. For example, Anaraky et. al. [3] showed that older and younger adults have different decision processes. While younger adults mostly rely on affect heuristics [44], older adults are more likely to be calculus-driven decision makers. However, to the best of our knowledge, no one has studied the differences between older and younger adults' decisions in the scope of dark pattern designs that nudge users towards disclosure. In this study, we investigate how dark pattern design interventions influence older adults' decisions differently than younger adults. Next, we turn to discussing the dark pattern design literature.

## 2.2 Dark Pattern Designs

Dark patterns in design refer to instances where designers exploit human desires and behaviors and implement functionality that will mislead them and have negative implications [28]. The term was first proposed by UX designer Harry Brignull in 2010, who defined dark patterns as: "a user interface that has been carefully crafted to trick users into doing things they might not otherwise do. "Brignull also notes that "Dark patterns are not mistakes. They are carefully crafted with a solid understanding of human psychology, and they do not have the user's interests in mind." [28]. Bösch et. al developed a categorization of privacy dark patterns that designers implement within their systems to exploit the users' privacy. These strategies include: maximize, publish, centralize, preserve, obscure, deny, violate, and fake [12].

In the context of this work, we focus on the *maximize* dark strategy by which designers aim to collect more data than is actually needed for the task [12]. Framing, defaults, and justification messages are means by which these designers maximize the amount of personal data collected [12, 33, 54]. We study these dark patterns in the context of a Facebook application—a context in which dark patterns are not uncommon. For example, in the infamous Cambridge Analytica case, the app designers used maximize dark pattern designs to maximize disclosure. The app used defaults to influence users' decisions to share personal information with rather unfavorable consequences. In our study, we test the effect of dark patterns in the context of a phototagging application on the Facebook platform where users can choose to tag themselves or their friends in their photos. The app applied maximize dark pattern designs to influence the user's decision in the form of choice framing, default settings, and justification messages. We explain each of these maximize dark patterns here.

### 2.2.1 Framing and Defaults

Framing and defaults are quintessential examples of the "maximize" dark pattern, in that they tend to increase compliance to disclosure requests made by the information system [4]. The framing effect describes the phenomenon that people are more likely to give consent to a request if it is presented with a positive framing ("Do
...") rather than a negative framing ("Do not ..."). Johnson et al. [30] and Lai and hui [43] independently studied the framing effect in the context of information privacy. At the end of an online health survey, Johnson et al. [30] asked their participants if they wanted to receive more health surveys. The wording of the choice option for participants in the positive framing condition was "Notify me about more health surveys", whereas those in a negative framing condition saw the choice wording as "Do not notify me about more health surveys". Lai and Hui
[43] studied framing in a newsletter sign-up scenario. Similarly, the wording for their positive framing condition was "Please send me Vortrex Newsletters and information" whereas the wording for the negative framing condition was "Please do not send me Vortrex Newsletters and information". Both Johnson and Lai and Hui found that participants are more likely to comply with the request if the request is presented with a positive framing rather than a negative framing.

Similar to framing, defaults are a form of dark pattern design strategy that can influence individuals' decisions [33, 69]. The default effect suggests that individuals are more likely to accept an option if that option is pre-selected by default [70]. This is evident in both Johnson et al. [30]'s health survey study and Lai and Hui [43]'s newsletter sign-up study, where sign-up ratio is highest if users are signed up by default (an opt-out default). Overall, both framing and defaults are referred to as tools of choice architecture [33] which can induce compliance to data disclosure requests made by the information system [4].

### 2.2.2 Justification Messages

To help users make a decision, system designers sometimes show them justification messages[1] providing additional information about the choice. These messages can inform users about the popularity of the product or service among other users [22] or its pros and cons [12, 24]. For example, Weinberger et al. [93] show that an unfavorable product rating adversely influences individuals' intention to purchase the product. Overall, these studies suggest that providing justifications supporting a request would motivate the audiences to comply to the request.

## 3 Research Framework

In the sections below, we justify information disclosure and privacy concerns as the two outcome variables of interest when examining the influence of dark-pattern designs. We also explain the 2x2x5 experimental design (where the conditions consist of varying defaults, framing, and justifications). Figure 1 shows our hypotheses via our research model. We investigate these hypotheses using a path model. In a path model, variables can function as both dependant variables (DVs) and independent variables (IVs). In our case, privacy concerns is a DV where we study the effects of dark pattern designs on it, and is an IV when we study how it predicts disclosure behaviors.

### 3.1 Disclosure Behavior

Oversharing information on social media can lead to negative consequences for users [65]. Therefore, social media platform users employ a wide range of privacy management strategies to manage their interpersonal boundaries, such as managing their relationship boundaries (e.g., by adding a new friend or unfriending someone) and their territorial boundaries (e.g., by tagging or untagging oneself or someone else in/from photos) [94].

---

[1] The term "framing" is used in the literature to denote several conceptually distinct interventions, and some studies apply the term "framing" to the type of justifications we use in this study [10]. In order to avoid confusion, we use the term "framing" for negated choice statements (i.e., "Do" vs. "Do not"; [30, 43]), and use the term "justification" to refer to the additional text accompanying the choice statement.

Existing literature regards tagging as a form of disclosure [95], since photo tagging can reveal the tagged person's online information (e.g. name, social media page) to a broader audience when the photo is shown on friends' timelines. Tagging other people in ones' photos is a contentious issue [95]: on the one hand, it can increase group cohesion and build social capital [56, 68], but on the other hand, it can lead to an interpersonal privacy violation if the others prefer not to be tagged [26, 81].

### 3.2 Privacy Concerns

The privacy literature has not reached a consensus about the relationship between privacy concerns and disclosure behaviors. On the one hand, many studies suggest a negative association between privacy concerns and disclosure behaviors [21, 34, 73]. The general argument in these studies is that a high concern for privacy motivates individuals to refrain from disclosing their data. On the other hand, many studies have found that despite their high privacy concerns, individuals freely give up their personal information—a phenomenon that is so prominent that it has been dubbed the "privacy paradox" [8]. The privacy paradox suggests that privacy concerns have little or no relationship with self-reported or observed disclosure behaviors [57, 79, 83]. For example, Tufekci [87] studied students' self-reported disclosure behaviors on social network sites. Her results show "little to no relationship" between online privacy concerns and disclosure on online social network sites. Despite these contradictory findings, we pose the following hypothesis reflecting the base expectation that privacy concerns are predictive of disclosure behaviors:

**H1**: *Individuals with higher privacy concerns are less likely to disclose.*

In the following two sub-sections, we build on our RQ1 by posing six hypotheses about the relationships between dark pattern design, disclosure behavior, and privacy concern. We then address our RQ2 by posing two hypotheses regarding the effect of age on disclosure behavior and privacy concerns.

### 3.3 Behavioral Effects of Dark Pattern Designs: Inducing Disclosure

In this section, we discuss the behavioral effects of framing, defaults, and justification messages. Particularly, we draw upon the "privacy dark patterns" literature to hypothesize how these design features are being used to increase disclosure.

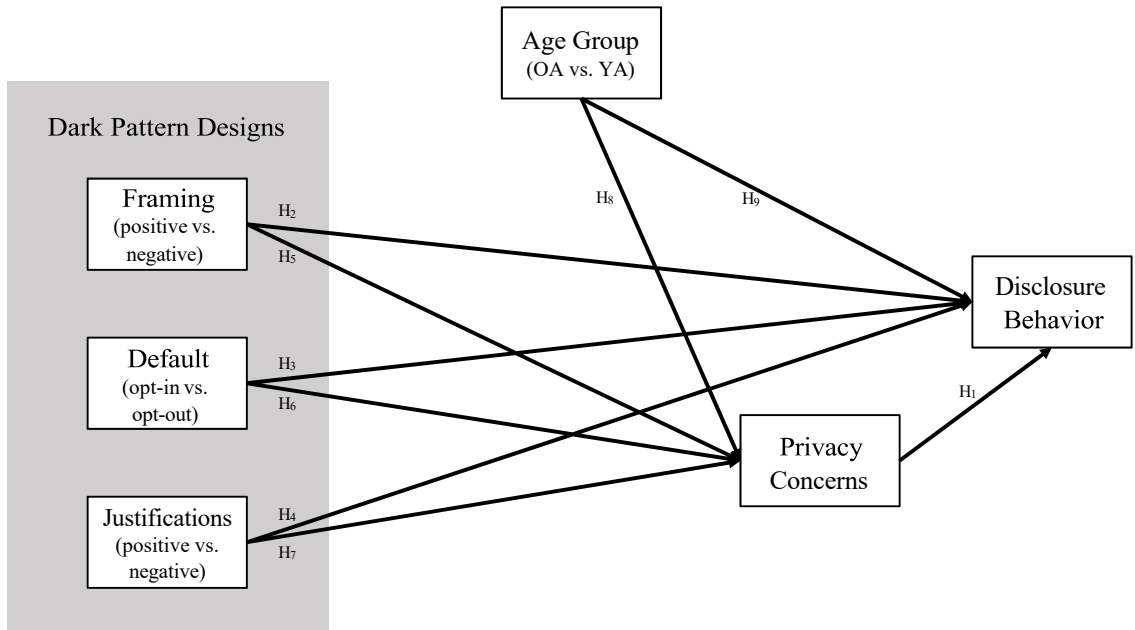

Fig. 1. Research model and the age-related effects of dark-pattern designs on privacy concerns and disclosure behavior (OA: older adults, YA: young adults).

### 3.3.1 Framing and Defaults

Framing and default effects have been extensively studied in the privacy domain. Johnson et al. [30] and Lai and Hui [43] independently found framing and default effects to have a significant impact on users' decisions. In both studies, a positive framing and an opt-out default setting increased the likelihood of users accepting the disclosure requests. In line with these findings, we study two conditions of framing (positive "tag me in the photos" vs. negative "do not tag me in the photos"), and two conditions of the default (opt-out vs. opt-in). We hypothesize the following:

**H2**: *A positive framing will increase disclosure.*
**H3**: *An opt-out default will increase disclosure.*

### 3.3.2 Justification Messages

Existing literature suggests that prompting individuals with a message about a product or service can influence their decisions in favor of the message content [27, 39, 45, 92]. Dark pattern designs sometimes use this feature to promote disclosure [12, 54]. For example, they show messages about a product's popularity in the form of reviews [17] which can sometimes be fake and misleading [80, 84, 91]. Hanson and Putler [27] showed participants a normative justification: an arbitrary integer ostensibly representing the number of downloads of a software program. Participants who saw a higher number were more likely to download the software for themselves. This is arguably due to a herding effect [6], where individuals follow the footsteps of the majority. In addition to such normative justifications, dark pattern designs sometimes use descriptive justification— pushing the benefits of the product—to promote product use or disclosure [12, 54]. For example, in the context of a travel advisor app, designers prompted users with a message of "by signing in, you can download over 300 cities, locations, and reviews to your phone" [12]. Reading about the benefits of the product is arguably a motivating message to sign in and use the app. Having higher quality reviews will increase the likelihood of one buying an online product [39], or talking about the effectiveness (pros) of treatment rather than the cons will increase the likelihood of accepting it [10]. In line with these findings, we study five conditions of justification messages (positive/negative normative, positive/negative rationale-based, none). We pose the following hypothesis:

**H4**: *Positive (as opposed to negative) justification messages will increase disclosure.*

### 3.4 Attitudinal Effects of Dark Pattern Designs: Inducing Concerns

The effects of framing, defaults, and justification messages on disclosure behavior are evident in the literature [1, 5, 27, 30, 33, 43]. The compliance-inducing effects of such design interventions meet the goal of dark pattern designs. However, research shows that the behavioral effects of these dark pattern strategies may not actually map to their attitudinal effects [7, 38, 41]. In this section, we examine the effects of dark pattern designs on individuals' state of privacy concerns. Privacy concerns are often measured as an individual trait, rather than a dynamic state [73]. However, this assumption may not be valid. Few studies consider privacy concerns as a dynamic state [67]. We further contribute to the literature by highlighting this perspective and presenting privacy concerns as a dynamic state which can change in response to different design strategies.

#### 3.4.1 Framing and Defaults

Online firms have an incentive to collect as much data about their users as they can. Data is an important asset in the industry [13, 29, 78]. For example, having data about customers' preferences and needs can help firms better target their advertisements to individuals likely to need their product and avoid the unnecessary costs of contacting consumers whose preferences do not match the product [29]. Therefore, some companies adopt dark-pattern design strategies to collect more user data [12].

Framing and default are compliance-inducing mechanisms [4] that are often used in dark pattern designs [25, 53, 54]. Dark pattern default options are designed to encourage sharing of personal information and to maximize online firms' collected data [12]. Likewise, dark pattern designs use positive framing in the choice statements to endorse disclosure [54]. Whereas the behavioral effect of such dark pattern designs are outlined in Section 3.3, here we note how these interventions can adversely influence users' attitudes towards the system [12, 47]. For instance, Knijnenburg and Kobsa [41] found that an opt-out default would increase perceived oversharing threats compared to an opt-in default. Therefore, we pose the following hypothesis:

**H5**: *A positive framing will increase privacy concerns.*

**H6**: *An opt-out default will increase privacy concerns.*

#### 3.4.2 Justification Messages

In the context of dark-pattern design, we argue that while a positive justification message might be intended to encourage individuals to use an app or increase disclosure, it could also increase users' privacy concerns, especially when they realize that the justification message is used as a dark-pattern strategy. Conversely, users may find a negative justification (i.e., cautioning them about the cons of the product) a sincere message [38] that is indicative of the developer's benevolence and/or integrity (two of the primary components of trust [55]). For example, communicating potential risks in a study report will increase its perceived trustworthiness [72]. Since trust and privacy concerns are inter-related [19], we argue that a negative justification will result in lower privacy concerns:

**H7**: *Negative (as opposed to positive) justification messages will decrease privacy concerns.*

Next, we explain the effects relating to age and our second research question.

### 3.5 Examining Differences between Older Adults and Young Adults

Literature suggests a positive association between age and privacy concerns, indicating higher concerns for older users [90]. However, the literature shows mixed findings in terms of the main effect of age on disclosure behaviors. While some studies do not report a significant relationship between age and disclosure [35], other studies show higher disclosure rates for older adults [71]. Despite these mixed findings, we pose the following hypotheses to investigate these effect:

**H8**: *Older adults will have higher levels of privacy concerns than young adults.*

**H9**: *Older adults will disclose more information than young adults.*

To the best of our knowledge, we are the first to study the difference between older and younger adults responses to the dark pattern designs.

# 4 Methods

## 4.1 Study Overview

To study our hypotheses, we used an existing dataset of a between subject experiment: a Facebook application that could purportedly automatically tag photos on Facebook. This application required users to have an active Facebook account with at least ten friends to be able to participate in the study. As a cover story, participants were told that the study would support the development of a Face-detection algorithm for a Facebook application that can automatically tag people in pictures and their task is to train this algorithm. Prior to logging on Facebook, the participants were asked to list the names of three of their friends with whom they have the most online interactions. After training phase, the app measured the repeated measures dependent variable: participants' tendency to use the tagging feature for tagging themselves (or each of these three friends) in their own (or in each of these three friends') photos. Figure 2 shows the experimental setup.

To make sure that participants understood the purported workings of the app, their first task was to test the readability of the following note: "This is a free application being developed by university researchers. It can automatically tag users or users' friends with high accuracy. Should the app make a mistake, users can still remove the tags." A short survey asked a number of comprehension questions about this note. Participants who answered these questions incorrectly had to read the note and answer the questions again. This procedure ensured that all participants clearly understood the context of the application.

Participants then entered the "training" phase of the application, in which they were asked to tag the people in four researcher-provided photos based on a key with the faces of individuals in the photo. This task was purportedly the main task for participants as it serves the cover story (i.e., training the application). Figure 3 shows the first training page. After training the application, participants started the "correction" phase of the study. In this phase, the app would display photos that were ostensibly tagged by the algorithm asking participants to correct any mistakes in the tags. All the photos were pre-tagged correctly so that participants do not have to make any corrections. The goal of the "correction" phase was to demonstrate the high accuracy of the tagging algorithm, thereby countering any potential fears about the possibility of the algorithm tag-ging users incorrectly. Training and correction phases were parts of the cover story, as the app told participants their main task would be training the algorithm. Participants did not know that the main task is their tagging decision. After this general training phase, the app provided participants with a scenario where they had to make a privacy decision (tag or not) relevant to each of their friends' names they listed before, where it measured the disclosure decision.

## 4.2 Dependent Variables: Tagging Decisions and Privacy Concerns

Participants finally entered the "decision" phase, where they were given a chance to use the tagging feature for themselves. In this stage, the app thanked participants for training the algorithm and, as a token of appreciation, offered them a chance to use the app for their own photos. The experimental manipulations in terms of default, framing, and justification messages (see "Manipulations" below) were applied here as betweensubjects manipulations. In a pre-questionnaire, a survey asked participants to enter the names of three Facebook friends that they regularly interact with; in the decision phase the app showed participants a decision page for each of these three friends. The decision page would claim that the algorithm had "identified" a) a number of previously unseen photos of the user on the friend's page, and b) a number of previously unseen photos of the friend on the user's page (in reality, the number of "identified" photos was a random number between five and fifteen). Participants were offered the choice to tag themselves, as well as the choice to tag their friend in these photos (see Figure 4). This resulted in six decisions (two tagging decisions for each of the three friends) per participant.

The results of a pilot study suggest that these six decisions hold credible ecological validity. In the pilot study, the disclosure scenario was more intrusive; instead of providing the tagging option only for three listed friends, the app inquired participants if they want to tag *all* of their friends in *all* of their own photos and tag themselves in *all* of their friends' photos. Out of 50 participants, no one agreed to use the tagging feature. Everyone rejecting the tagging across different framing, defaults, and justification manipulations means that participants perceived the app as real. After this pilot, the app designers adopted a less intrusive scenario to proceed with their study.

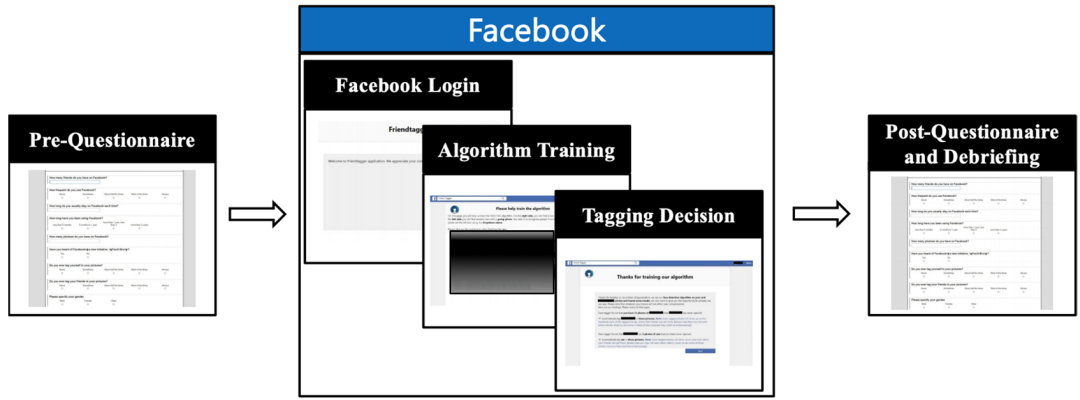

Fig. 2. The Experimental Setup. This figure is anonymized.

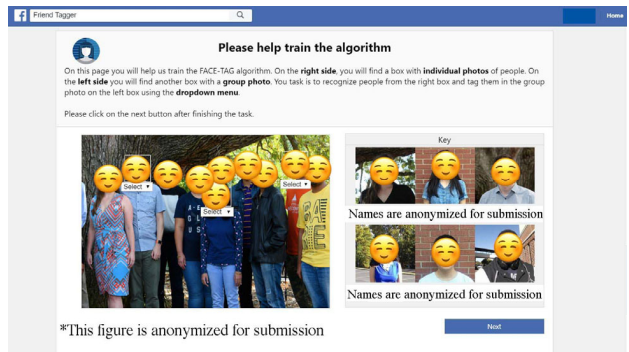

Fig. 3. One of the training pages where participants could tag individuals in the photo on the left side of the screen based on a key on the right side of the screen. They were told that they are "training" the algorithm here. This figure is anonymized.

Privacy concerns is the second dependent variable of our interest. The dataset includes a reduced version of the IUIPC [48] dimension of general concerns with 3 items. This reduced version is validated and used in the previous studies [40]. We used the sum score of this scale for measuring privacy concerns (Cronbach's $\alpha = 0.790$). All the items were measured on a 5-point agreeableness Likert scale:

– Compared to others, I am more sensitive about the way online companies handle my personal information.
– To me, it is the most important thing to keep my privacy intact from online companies.
– I am concerned about threats to my personal privacy today.

We standardized the scale (grand mean = 0, SD = 1) in our analyses.

### 4.3 Independent Variables: Manipulations

Similar to most existing studies on framing and default effects, this experiment combined a default manipulation (accept vs. reject) with a framing manipulation (positive vs. negative). Table 1 shows this 2x2 design.

In addition to framing and defaults, the app involves a justification manipulation which adds a "justification message" to the decision scenario. Literature studies two common types of justification messages: normative [27] and rationale-based [66] justifications. For the sake of robustness, the experiment includes both types of justifications, each with both a positive and a negative valence. Therefore, the justifications manipulation consists of two normative justifications (positive—showing high popularity for the app, and negative—showing low popularity for the app), two rationale-based justifications (positive—discussing pros of using the app, and negative—discussing cons of using the app), and a condition without any justification (as a neutral baseline):

– **Negative descriptive normative justification:** Note: 3% of our study participants use the tagging feature.
– **Positive descriptive normative justification:** Note: 97% of our study participants use the tagging feature.
– **Negative rationale-based justification:** Note: Auto-tagged photos will show up on the Facebook walls of the tagged friends, where their friends can see them. Beware that they may not want others (parents, boss) to see some of these photos, because they could be embarrassing!

Table 1. A representation of framing and default conditions

| Presentation of the choice option | Framing | Default |
|---|---|---|
| ☒ Automatically tag me in the photos. | Positive | Opt-out |
| ☐ Automatically tag me in the photos. | Positive | Opt-in |
| ☒ Do NOT automatically tag me in the photos | Negative | Opt-in |
| ☐ Do NOT automatically tag me in the photos | Negative | Opt-out |

- **Positive rationale-based justification:** Note: Auto-tagged photos will show up on the Facebook walls of the tagged friends, where their friends can see them. This will strengthen your friendship and let your friends relive the good times they had with you!
- **None:** No justification given. (This condition was treated as a baseline control for the model)

All of the experimental manipulations were betweensubjects. An example condition is shown in Figure 4.

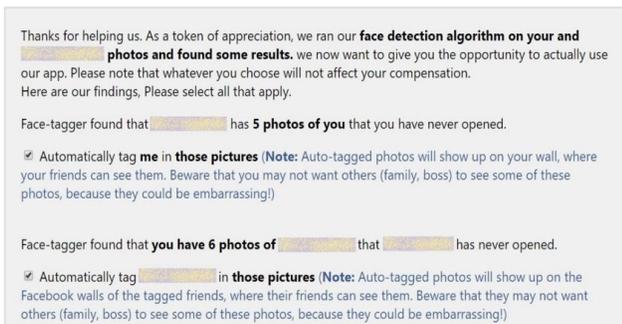

Fig. 4. The Decision Page. This is an example for positive framing and opt-out default conditions. In addition, there is a positive rationale-based justification provided in blue, between the parenthesis.

### 4.4 Participant Recruitment

This study was approved by an institutional review board. After excluding those who failed an attentioncheck question, we were left with 44 older adults and 169 young adults. These participants were recruited through MTurk and Figure-eight crowd-sourcing platforms. There were not significant demographic differences between the participants across these platforms. Each participant received $1.30 (U.S. Dollars) as their participation incentive. The app required participants to have an active Facebook account with at least ten friends to participate in the study. Participants were debriefed after the study and were informed that the app was not real and that it did not actually tag any of their or their friends' photos. As is customary in studies that involve deception, they were given the option to be removed from the data without influencing their incentives. No participants chose to do so.

### 4.5 Data Analysis Approach

Our first dependant variable (decision) was a binary variable with "accept to tag" coded as 1 and "reject to tag" coded as 0. Each participant responded to six disclosure scenarios: three scenarios on whether they wanted to tag themselves in each of their three friends' photos and three scenarios on whether they wanted to tag each of their three friends in their own photos. Therefore, we constructed a multilevel path model with a random intercept to account for repeated measures per participant and a binary dependent variable. Our path model enabled us to treat privacy concerns as both an independent variable (by regressing tagging decision on it) and a dependent variable (by regressing it on study manipulations). The framing and default manipulations were dummy-coded where positive framing or opt-out defaults were coded as "0.5" and negative framing or opt-in defaults were coded as "−0.5". To analyze justification messages, we first conducted an overall chi-square omnibus test to study their overall effect among the 5 conditions. Then we ran planned contrasts, including a contrast testing the effect of justification valence (positive justifications vs. negative justifications) to study H4 and H7. The other contrasts tested the effect of any justification (no vs. any), the effect of the type of justification (rationale-based vs. normative), and the interaction between justification type and valence). Our analyses were carried out in Mplus v7.4. As the sample was imbalanced (44 older adults and 169 young adults), we used MLR, a maximum likelihood estimation with robust standard errors in our analyses [59].

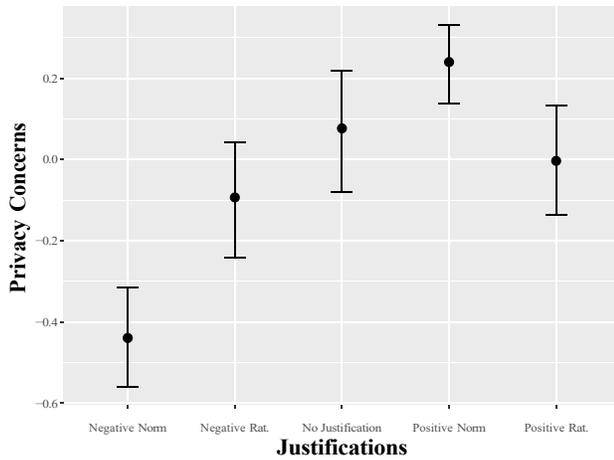

Fig. 5. The y-axis is standardized sum-score of privacy concerns. The graph shows that negative justifications lead to lower levels of privacy concerns.

# 5 Results

## 5.1 Sample Characteristics

Across the older adult participants, there were 15 females and 29 males. Their average age was 68.8 years (min = 65, max = 77, sd = 3.23). The young adult sample had 87 males, 71 females, and 4 individuals who did not self-identify as female or male. Their average age was 20.30 (min = 18, max = 25, sd = 2.18). The dataset also measured participants' Facebook usage frequency on a 9-point scale from "Never" coded as 0 to "Almost constantly" coded as 9 and the time they spend on Facebook in each session on a 5-point scale from "A few minutes" coded as 1 to "Several hours" coded as 6. On average, older adults used the Facebook platform more frequently (mean = 5.708, sd = 1.519) compared to younger adults (mean = 5.402, sd = 2.055). However, both age groups spent an average time of 30 minutes on Facebook (older adults: mean = 2.054, sd = 1,194 and young adults: mean = 2.119, sd = 1.275). In the following, we present our analyses with regards to our hypotheses. Table 2 summarizes the hypothesis tests.

## 5.2 Effects on Tagging Decision

In contrast to H1 and in line with the privacy paradox, we did not find any significant relationship between privacy concerns and disclosure decision ($p>.05$). Therefore, H1 is rejected. However, we found support for H2 and H3 suggesting both framing and default significantly influence tagging decisions. Users who see the positively framed option of "Tag me in the photos" were 31.9% more likely [2] to use the tagging feature compared to those who see the negatively framed option of "Do not tag me in the photos" ($p < .001$). Furthermore, those users who were being tagged by default were 19.4% more likely to use the tagging feature ($p<.01$). In addition, in contrast to H4, the various justifications did not influence tagging decision differently ($\chi^2(4) = 0.823, p > 0.05$). Finally, in line with what we hypothesized in H9, older adults were 19.8% more likely to use the tagging feature compared to young adults ($p<.05$).

## 5.3 Effects on Privacy Concerns

In contrast to H5 and H6, we did not find any significant effects of framing and default manipulations on privacy concerns ($p > .05$). However, we did find a significant main effect of justifications on privacy concerns ($\chi^2(4) = 9.827, p < 0.05$). Subsequent planned contrast tests revealed a significant effect of justification valence ($p < 0.05$), supporting H7 and suggesting that negative justification messages lower users' privacy concerns compared to positive justifications by 0.132 standard deviations. However, this effect was negated by a marginal interaction between justification type and valence, meaning that the difference between positive and negative justifications mostly hold for normative justifications (see Figure 5). Finally, in contrast to H8, older and younger adults did not have significant differences in terms of their overall level of privacy concerns ($p > 0.05$).

Our results show that in contrast to our H8, older and younger adults have the same levels of privacy concerns. Furthermore, while having the same levels of concern, older adults are more open to disclosure. We further unpack the effects of age by studying moderation effects of age and dark pattern designs. We therefore run a saturated path model and report it below. Figure 6 and Table A1 summarize our results..

---

[2] To present the results in a comprehensive manner, we convert the log odds-ratio to percentages. For example, the effect of framing on disclosure is 0.277 (table A1), which is a log odds-ratio. Therefore, disclosure in the positive framing group is $e^{0.277} = 1.319$ times higher than in the negative framing group, i.e., a 31.9% difference in the odds of disclosure

Table 2. An overview of the study's results regarding the hypotheses

| Hypothesis | Support |
|---|---|
| H1: High privacy concerns –> Low disclosure | Not supported |
| H2: Positive framing (vs. negative) –> High disclosure | Supported |
| H3: Opt-out default (vs. opt-in) –> High disclosure | Supported |
| H4: Positive (vs. negative) justifications –> High disclosure | Not supported |
| H5: Positive (vs. negative) framing –> High privacy concerns | Not supported |
| H6: Opt-out (vs. opt-in) default –> High privacy concerns | Not supported |
| H7: Negative (vs. positive) justifications –> Low privacy concerns | Supported |
| H8: Older adults (vs. younger adults) –> High concerns | Not supported |
| H9: Older adults (vs. younger adults) –> High disclosure | Supported |

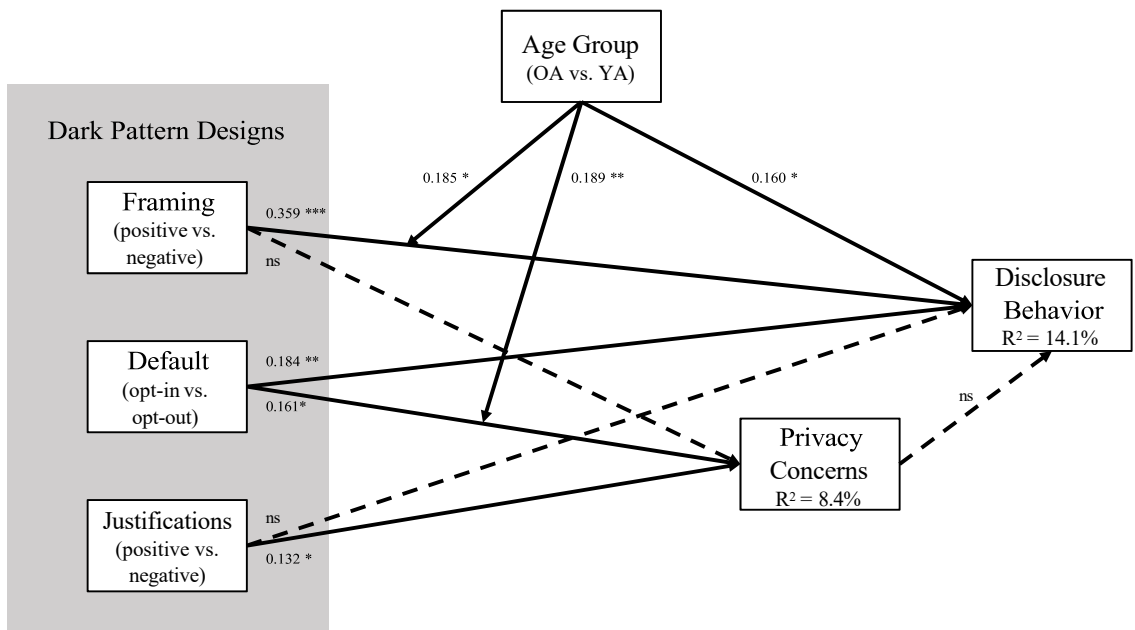

Fig. 6. The results of our saturated path model including all of the significant findings (ns: not significant hypotheses, * $p < .05$, ** $p < .01$, *** $p < .001$)

### 5.4 Moderating Effects of Age Group on Tagging Decision

We found age group to moderate the effect of framing on tagging decision; the effect of framing was stronger for older adults than for young adults: older adults who were exposed to a positively framed option were 53.1% more likely to use the tagging feature while young adults who were exposed to a positively framed option were only 33.1% more likely to use the tagging feature ($p < .05$). Figure 7 depicts this effect. We found a similar moderation effect for defaults suggesting that older adults were more likely to keep the default option and use the tagging feature compared to young adults. However, this effect did not reach significance ($p = .062$).

Similarly, age did not moderate the effect of justifications on the tagging decision ($\chi^2(4) = 4.822, p > 0.05$). Lastly, while the effect of privacy concerns on the tagging decision was stronger for older adults than for young adults, this effect was not significant ($p > .05$).

### 5.5 Moderating Effects of Age Group on Privacy Concern

The effect of defaults on privacy concerns were moderated by age group: an opt-out default increased older adults privacy concerns by 0.255 times standard deviation compared to a negative default, while an opt-out default increased young adults' privacy concerns by only 0.066 times standard deviation compared to a negative default ($p < 0.01$). Figure 8 depicts this effect and suggests that an opt-out default only significantly increases the privacy concerns for older adults. Age group did not moderate the effects of any other variable on privacy concerns ($p > 0.05$).

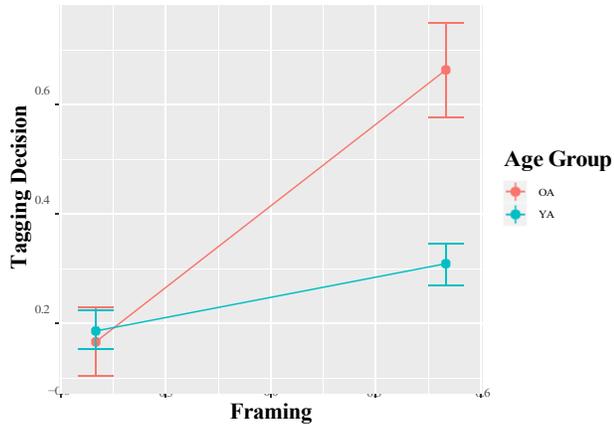

Fig. 7. The y-axis shows the actual tagging decision with zero as reject and one as accept. The graph shows that a positive framing is a more effective dark pattern strategy for older adults in terms of inducing disclosure.

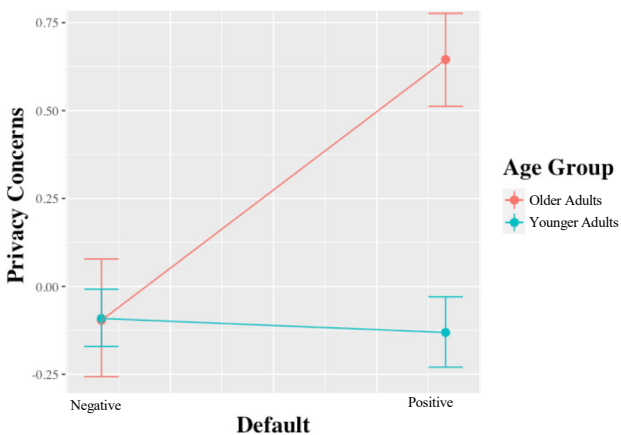

Fig. 8. An opt-out default can increase older adults' concerns for privacy

# 6 Discussion

In this section, we first discuss the main effects of dark pattern design strategies on users' disclosures and privacy concerns (RQ1). We then discuss how dark pattern designs influence older adults differently compared to younger adults (RQ2). Finally, we conclude the discussion section by presenting the design implications of this work.

## 6.1 The Impact of Dark Pattern Designs on Disclosure Behaviors

In this study, we did not find support for H1 and therefore did not find privacy concerns significantly influence the tagging decision. This finding confirms the privacy paradox theory that individuals' disclosure behaviors are not necessarily in-line with their self-reported privacy concerns [8]. However, we should consider that the effect of privacy concerns on disclosure was not studied in isolation. We had framing and default nudges present in the decision scenario which, indeed, significantly influenced users' decisions (H2, H3). It is possible that concern for privacy was not a determining factor for users in the presence of such dark pattern design interventions. As an implication of our RQ1, future studies should focus more on dark pattern designs, even more so than privacy concerns, as dark pattern designs have a stronger effect on users' disclosure behaviors. We see that even raising privacy concern does not translate to privacy protective behaviors.

## 6.2 The Impact of Dark Pattern Designs on Privacy Concern

Despite many privacy frameworks depicting privacy concerns as a dynamic and very fluid concept [2, 61, 63], the majority of the privacy literature has studied privacy concerns as a static variable predicting disclosure [73, 97]. We studied privacy concerns as a dynamic result of design patterns. Our results suggest that concern for privacy is not necessarily a static trait, and rather can change in response to the design patterns. Specifically, negative justifications can decrease users' privacy concerns. A possible explanation is that showing a negative justification makes users feel that the app is more sincere. This, in turn, lowers levels of privacy concern. For example, showing a negative normative justification is an indication of low popularity of a product and is an uncommon practice and induces lowers levels of privacy concern.

### 6.3 Dark Pattern Designs May Disproportionately and Negatively Impact Older Adults

In terms of RQ2, we studied the the difference between older and younger adults privacy concerns and disclosures. In addition, we studied the moderating effects of age. While older and younger adults had similar privacy concerns, our results show that an opt-out default dark pattern alerts older adult users and makes them privacy-cautious. Using opt-out defaults is one of the most common means of data collection by firms [30]. It is possible that our older adult participants had more experience and familiarity with the default mechanisms and, therefore, an opt-out default led them to be more privacy-cautious.

With regards to the effects of age on tagging decisions, we found that older adults were more likely to use the tagging feature. This is in-line with some previous findings in the literature on older adults disclosing more data [71]. In addition to this main effect of age on decision, we found that the framing effect is a stronger nudge in pushing older adults to use the tagging feature when compared to younger adults. Likewise, we found opt-out defaults to be a stronger nudge for older adult participants; however, the moderation effect with defaults did not reach the significant thresholds ($p = 0.062$). These effects may be explained by the literature on loss-aversion. Losses are weighed more heavily than gains and so individuals put forth more effort to avoid losses than to acquire gains [88, 89]. An opt-out default can trigger an instant endowment for the user, where the tagging disclosure is seen as something they have [20, 51]. Therefore, changing the default is being perceived as a loss and individuals are more likely to keep the default option [36, 37]. Likewise, a positive framing endows individuals with the benefits of disclosure, but foregoing disclosure is perceived as a loss [30, 32]. This is further supported by the psychology literature which suggests that older adults are generally more loss-adverse than young adults [16, 18, 31, 52]. Scholars have found that older adults are willing to take more risks [52] or exert more effort [16] to avoid a loss, in comparison with young adults. Therefore, our framing and default manipulations triggered a loss-aversion process which influenced older adults more than younger adults.

### 6.4 Implications for Design

Our study has several design implications. A key finding is that while using opt-out defaults increases older adults' privacy concerns, it still ends up increasing their disclosure levels. This goes counter to the common perception about older adults having low privacy awareness, since they identify an opt-out dark pattern design—even more so than younger adults—and become privacy-couscous. However, these dark pattern interventions had stronger behavioral effects than any heightened privacy concerns. Therefore, instead of efforts to make individuals privacy-couscous and increase individuals' privacy concerns, hoping for them to take privacy protective measures, it may be more effective to focus on how to counter dark design patterns. This might even include developing policies that discourage or regulate the use of dark design patterns.

We also found that older adults may be more amenable to framing and default nudges due to their loss-aversive nature. This result is a call to technology developers to be mindful of their older adult audiences and take on the ethical responsibility of creating technologies that avoid such nudges. In fact, prior research suggests that older adults may choose not to use technology as a result of high privacy concerns [42]. While the opt-out default increased disclosure in this study, it is conceivable that having to make a plethora of lossaversive decisions could push privacy concerns beyond a threshold where older adults decide to stop using technology altogether. Further research is needed to investigate this, but in the meantime, product designers should be conscientious towards their older adult users and not increase their concerns.

Furthermore, while it may seem counter intuitive, if product designers are honest about the negative aspects of their products, especially the low adoption of their features, it may actually alleviate concerns. Our negative justifications manipulation proved to reduce privacy concerns. Being honest seems to be the best policy for gaining consumer confidence. Finally, Designers can use various methods such as graph-based models to assess the potential impact of their design choices on user behavior and ensure that they are not inadvertently creating dark patterns [74, 76, 77].

# 7 Limitations and Future Work

While we were able to study both older and younger adults to gain insight into how dark patterns differently affect these age groups, this was an initial exploration with a non representative sample. Future research should study this phenomenon with a bigger sample of participants that are balanced to be representative. Also, since privacy is a culturally-shaped construct, investigating attitudes in other cultures and countries would broaden our understanding beyond the United States. Furthermore, we studied disclosure behaviors in the context of tagging on social media. The effect of these patterns may vary in different contexts. Thus, future research should consider privacy decisions made in context of other domains such as for e-commerce or healthcare services.

# 8 Conclusion

We studied the attitudinal and behavioral impact of dark pattern designs on older and younger adults. While an individual's levels of privacy concerns may change in response to these design strategies, the behavioral effects of such strategies are dominant and individuals still end up disclosing their data despite heightened concerns. Furthermore, while older adults respond with more concern to some of these dark pattern designs than young adults, they are actually more vulnerable to such design strategies, perhaps due to a loss-aversive nature. Therefore, policy designers and technology developers should become familiar with the unique privacy attitudes and behaviors of older adults when it comes to disclosure. The solution may be a combination of identifying technology designs that counter the effects of dark patterns, as well as establishing rules and regulations around their use. Until that happens, older adults may continue to be disproportionately affected by dark pattern designs.

# References

[1] Idris Adjerid, Alessandro Acquisti, Laura Brandimarte, and George Loewenstein. Sleights of privacy: Framing, disclosures, and the limits of transparency. In *Proceedings of the ninth symposium on usable privacy and security*, pages 1–11, 2013.

[2] Irwin Altman. The environment and social behavior: privacy, personal space, territory, and crowding. 1975.

[3] Reza Ghaiumy Anaraky, Kaileigh A. Byrne, Pamela J. Wisniewski, Xinru Page, and Bart P. Knijnenburg. To disclose or not to disclose: Examining the privacy decision-making processes of older vs. younger adults. In *To Appear Proceedings of the 2021 CHI Conference on Human Factors in Computing Systems*. ACM, 2021.

[4] Reza Ghaiumy Anaraky, Bart P Knijnenburg, and Marten Risius. Exacerbating mindless compliance: The danger of justifications during privacy decision making in the context of facebook applications. *AIS Transactions on HumanComputer Interaction*, 12(2):70–95, 2020.

[5] Reza Ghaiumy Anaraky, Tahereh Nabizadeh, Bart P Knijnenburg, and Marten Risius. Reducing default and framing effects in privacy decision-making. *Proceedings of the Special Interest Group On Humancomputer Interaction*, 2018.

[6] Solomon E Asch. Studies of independence and conformity: I. a minority of one against a unanimous majority. *Psychological monographs: General and applied*, 70(9):1, 1956.

[7] Paritosh Bahirat, Martijn C. Willemsen, Yangyang He, Qizhang Sun, and Bart P. Knijnenburg. Overlooking context: How do defaults and framing reduce deliberation in smart home privacy decision-making? In *To Appear Proceedings of the 2021 CHI Conference on Human Factors in Computing Systems*. ACM, 2021.

[8] Susan B Barnes. A privacy paradox: Social networking in the united states. *First Monday*, 2006.

[9] Andrew Besmer and Heather Richter Lipford. Moving beyond untagging: photo privacy in a tagged world. In *Proceedings of the SIGCHI Conference on Human Factors in Computing Systems*, pages 1563–1572, 2010.

[10] Cabral A Bigman, Joseph N Cappella, and Robert C Hornik. Effective or ineffective: Attribute framing and the human papillomavirus (hpv) vaccine. *Patient education and counseling*, 81:S70–S76, 2010.

[11] Jeremy Birnholtz and McKenzie Jones-Rounds. Independence and interaction: understanding seniors' privacy and awareness needs for aging in place. In *Proceedings of the SIGCHI Conference on Human Factors in Computing Systems*, pages 143–152, 2010.

[12] Christoph Bösch, Benjamin Erb, Frank Kargl, Henning Kopp, and Stefan Pfattheicher. Tales from the dark side: Privacy dark strategies and privacy dark patterns. *Proceedings on Privacy Enhancing Technologies*, 2016(4):237–254, 2016.

[13] Marc Bourreau, Alexandre De Streel, and Inge Graef. Big data and competition policy: Market power, personalised pricing and advertising. *Personalised Pricing and Advertising (February 16, 2017)*, 2017.

[14] Petter Bae Brandtzæg, Marika Lüders, and Jan Håvard Skjetne. Too many facebook "friends"? content sharing and sociability versus the need for privacy in social network sites. *Intl. Journal of Human–Computer Interaction*, 26(1112):1006–1030, 2010.

[15] US Census Bureau. Historical Living Arrangements of Adults. Section: Government.

[16] Kaileigh A Byrne and Reza Ghaiumy Anaraky. Strive to win or not to lose? age-related differences in framing effects on effort-based decision-making. *The Journals of Gerontology:*

[17] Chrysanthos Dellarocas, Xiaoquan Michael Zhang, and Neveen F Awad. Exploring the value of online product reviews in forecasting sales: The case of motion pictures. *Journal of Interactive marketing*, 21(4):23–45, 2007.

[18] Miriam K Depping and Alexandra M Freund. Normal aging and decision making: The role of motivation. *Human Development*, 54(6):349–367, 2011.

[19] Tamara Dinev, Allen R McConnell, and H Jeff Smith. Research commentary—informing privacy research through information systems, psychology, and behavioral economics: thinking outside the "apco" box. *Information Systems Research*, 26(4):639–655, 2015.

[20] Isaac Dinner, Eric J Johnson, Daniel G Goldstein, and Kaiya Liu. Partitioning default effects: why people choose not to choose. *Journal of Experimental Psychology: Applied*, 17(4):332, 2011.

[21] Mary Ann Eastlick, Sherry L Lotz, and Patricia Warrington. Understanding online b-to-c relationships: An integrated model of privacy concerns, trust, and commitment. *Journal of business research*, 59(8):877–886, 2006.

[22] Kelwin Fernandes, Pedro Vinagre, and Paulo Cortez. A proactive intelligent decision support system for predicting the popularity of online news. In *Portuguese Conference on Artificial Intelligence*, pages 535–546. Springer, 2015.

[23] Alisa Frik, Julia Bernd, Noura Alomar, and Serge Egelman. A qualitative model of older adults' contextual decisionmaking about information sharing. In *Workshop on the Economics of Information Security (WEIS 2020)*, 2020.

[24] Eyal Gamliel and Eyal Peer. Attribute framing affects the perceived fairness of health care allocation principles. *Judgment and Decision Making*, 5(1):11, 2010.

[25] Colin M Gray, Yubo Kou, Bryan Battles, Joseph Hoggatt, and Austin L Toombs. The dark (patterns) side of ux design. In *Proceedings of the 2018 CHI Conference on Human Factors in Computing Systems*, pages 1–14, 2018.

[26] Keith N Hampton, Lauren Sessions Goulet, Cameron Marlow, and Lee Rainie. Why most facebook users get more than they give. *Pew Internet & American Life Project*, 3(2012):1–40, 2012.

[27] Ward A Hanson and Daniel S Putler. Hits and misses: Herd behavior and online product popularity. *Marketing letters*, 7(4):297–305, 1996.

[28] Jeremy Rosenberg Harry Brignull, Marc Miquel and James Offer. Dark patterns user interfaces designed to trick people., 2015.

[29] Ganesh Iyer, David Soberman, and J Miguel Villas-Boas. The targeting of advertising. *Marketing Science*, 24(3):461–476, 2005.

[30] Eric J Johnson, Steven Bellman, and Gerald L Lohse. Defaults, framing and privacy: Why opting in-opting out 1. *Marketing letters*, 13(1):5–15, 2002.

[31] Eric J Johnson, Simon Gächter, and Andreas Herrmann. Exploring the nature of loss aversion. 2006.

[32] Eric J Johnson, John Hershey, Jacqueline Meszaros, and Howard Kunreuther. Framing, probability distortions, and insurance decisions. *Journal of risk and uncertainty*, 7(1):35–51, 1993.

[33] Eric J Johnson, Suzanne B Shu, Benedict GC Dellaert, Craig Fox, Daniel G Goldstein, Gerald Häubl, Richard P Larrick, John W Payne, Ellen Peters, David Schkade, et al. Beyond nudges: Tools of a choice architecture. *Marketing Letters*, 23(2):487–504, 2012.

[34] Adam N Joinson, Ulf-Dietrich Reips, Tom Buchanan, and Carina B Paine Schofield. Privacy, trust, and self-disclosure online. *Human–Computer Interaction*, 25(1):1–24, 2010.

[35] Sidney M Jourard. Age trends in self-disclosure. *MerrillPalmer Quarterly of Behavior and Development*, 7(3):191–197, 1961.

[36] Daniel Kahneman, Jack L Knetsch, and Richard H Thaler. Anomalies: The endowment effect, loss aversion, and status quo bias. *Journal of Economic perspectives*, 5(1):193–206, 1991.

[37] Daniel Kahneman, Jack L Knetsch, and Richard H Thaler. The endowment effect, loss aversion, and status quo bias: Anomalies. *Journal of Economic perspectives*, 5(1):193–206, 1991.

[38] Gideon Keren. Framing, intentions, and trust–choice incompatibility. *Organizational Behavior and Human Decision Processes*, 103(2):238–255, 2007.

[39] HyeKyoung Kim and Jihoon Song. The quality of word-ofmouth in the online shopping mall. *Journal of Research in Interactive Marketing*, 2010.

[40] Bart P Knijnenburg and Alfred Kobsa. Making decisions about privacy: information disclosure in context-aware recommender systems. *ACM Transactions on Interactive Intelligent Systems (TiiS)*, 3(3):1–23, 2013.

[41] Bart Piet Knijnenburg and Alfred Kobsa. Increasing sharing tendency without reducing satisfaction: Finding the best privacy-settings user interface for social networks. In *ICIS*, 2014.

[42] Bran Knowles and Vicki L Hanson. The wisdom of older technology (non) users. *Communications of the ACM*, 61(3):72–77, 2018.

[43] Yee-Lin Lai and Kai-Lung Hui. Internet opt-in and optout: investigating the roles of frames, defaults and privacy concerns. In *Proceedings of the 2006 ACM SIGMIS CPR conference on computer personnel research: Forty four years of computer personnel research: achievements, challenges & the future*, pages 253–263, 2006.

[44] Roy J Lewicki and Chad Brinsfield. Framing trust: trust as a heuristic. *Framing matters: Perspectives on negotiation research and practice in communication*, pages 110–135, 2011.

[45] Chia-Ying Li. Persuasive messages on information system acceptance: A theoretical extension of elaboration likelihood model and social influence theory. *Computers in Human Behavior*, 29(1):264–275, 2013.

[46] Lesa Lorenzen-Huber, Mary Boutain, L Jean Camp, Kalpana Shankar, and Kay H Connelly. Privacy, technology, and aging: A proposed framework. *Ageing International*, 36(2):232–252, 2011.

[47] Aditi M. Bhoot, Mayuri A. Shinde, and Wricha P. Mishra. Towards the identification of dark patterns: An analysis based on end-user reactions. In *IndiaHCI'20: Proceedings of the 11th Indian Conference on Human-Computer Interaction*, pages 24–33, 2020.

[48] Naresh K Malhotra, Sung S Kim, and James Agarwal. Internet users' information privacy concerns (iuipc): The construct, the scale, and a causal model. *Information systems*

*research*, 15(4):336–355, 2004.

[49] Aqdas Malik, Dhir Amandeep, Marko Nieminen, et al. Facebook photo tagging culture and practices among digital natives. In *CCGIDIS 2015-Fifth International Symposium on Communicability, Computer Graphics and Innovative Design for Interactive Systems*. Blue Herons Editions, 2015.

[50] Mehdi Marani, Morteza Soltani, Mina Bahadori, Masoumeh Soleimani, and Atajahangir Moshayedi. The role of biometric in banking: A review. *EAI Endorsed Transactions on AI and Robotics*, 2(1), 2023.

[51] Isabel Marcin and Andreas Nicklisch. Testing the endowment effect for default rules. *Review of Law & Economics*, 13(2), 2017.

[52] Mara Mather, Nina Mazar, Marissa A Gorlick, Nichole R Lighthall, Jessica Burgeno, Andrej Schoeke, and Dan Ariely. Risk preferences and aging: The "certainty effect" in older adults' decision making. *Psychology and aging*, 27(4):801, 2012.

[53] Arunesh Mathur, Gunes Acar, Michael J Friedman, Elena Lucherini, Jonathan Mayer, Marshini Chetty, and Arvind Narayanan. Dark patterns at scale: Findings from a crawl of 11k shopping websites. *Proceedings of the ACM on HumanComputer Interaction*, 3(CSCW):1–32, 2019.

[54] Arunesh Mathur, Jonathan Mayer, and Mihir Kshirsagar. What makes a dark pattern... dark? design attributes, normative considerations, and measurement methods. *arXiv preprint arXiv:2101.04843*, 2021.

[55] D Harrison McKnight, Vivek Choudhury, and Charles Kacmar. Developing and validating trust measures for ecommerce: An integrative typology. *Information systems research*, 13(3):334–359, 2002.

[56] Andrew L Mendelson and Zizi Papacharissi. Look at us: Collective narcissism in college student facebook photo galleries. *The networked self: Identity, community and culture on social network sites*, 1974:1–37, 2010.

[57] Miriam J Metzger. Effects of site, vendor, and consumer characteristics on web site trust and disclosure. *Communication Research*, 33(3):155–179, 2006.

[58] Mehrdad Mohammadi, Mahrou Samizadeh, Shaheen Pouya, and Raheleh Arabahmadi. Examining the mediating effect of knowledge management on the relationship between organizational culture and organizational performance. *Journal of Soft Computing and Decision Analytics*, 1(1):63–79, 2023.

[59] Bengt Muthén and Linda Muthén. *Mplus*. Chapman and Hall/CRC, 2017.

[60] Barbara M Newman and Philip R Newman. *Development through life: A psychosocial approach*. Cengage Learning, 2017.

[61] Helen Nissenbaum. Privacy as contextual integrity. *Wash. L. Rev.*, 79:119, 2004.

[62] Xinru Page and Marco Marabelli. Changes in social media behavior during life periods of uncertainty. In *Proceedings of the International AAAI Conference on Web and Social Media*, volume 11, 2017.

[63] Leysia Palen and Paul Dourish. Unpacking" privacy" for a networked world. In *Proceedings of the SIGCHI conference on Human factors in computing systems*, pages 129–136, 2003.

[64] Yong Jin Park. Digital literacy and privacy behavior online. *Communication Research*, 40(2):215–236, 2013.

[65] Chris Rose. The security implications of ubiquitous social media. *International Journal of Management & Information Systems (IJMIS)*, 15(1), 2011.

[66] Alexander J Rothman and Peter Salovey. Shaping perceptions to motivate healthy behavior: the role of message framing. *Psychological bulletin*, 121(1):3, 1997.

[67] Matthew Rueben, Frank J Bernieri, Cindy M Grimm, and William D Smart. Framing effects on privacy concerns about a home telepresence robot. In *Proceedings of the 2017 ACM/IEEE International Conference on Human-Robot Interaction*, pages 435–444, 2017.

[68] Paul Russo and Oded Nov. Photo tagging over time: A longitudinal study of the role of attention, network density, and motivations. In *Proceedings of the International AAAI Conference on Web and Social Media*, volume 4, 2010.

[69] Sonam Samat and Alessandro Acquisti. Format vs. content: the impact of risk and presentation on disclosure decisions. In *Thirteenth Symposium on Usable Privacy and Security ({SOUPS} 2017)*, pages 377–384, 2017.

[70] William Samuelson and Richard Zeckhauser. Status quo bias in decision making. *Journal of risk and uncertainty*, 1(1):7–59, 1988.

[71] Hu Shuijing and Jiang Tao. An empirical study on digital privacy risk of senior citizens. In *2017 International Conference on Robots & Intelligent System (ICRIS)*, pages 19–24. IEEE, 2017.

[72] Michael Siegrist and George Cvetkovich. Better negative than positive? evidence of a bias for negative information about possible health dangers. *Risk analysis*, 21(1):199–206, 2001.

[73] H Jeff Smith, Tamara Dinev, and Heng Xu. Information privacy research: an interdisciplinary review. *MIS quarterly*, pages 989–1015, 2011.

[74] M Soleimani, F Mahmudi, and MH Naderi. Some results on the maximal graph of commutative rings. *Advanced Studies: Euro-Tbilisi Mathematical Journal*, 16(supp1):21–26, 2023.

[75] Masoumeh Soleimani, Zahra Forouzanfar, Morteza Soltani, and Majid Jafari Harandi. Imbalanced multiclass medical data classification based on learning automata and neural network. *EAI Endorsed Transactions on AI and Robotics*, 2, 2023.

[76] Masoumeh Soleimani, Fatemeh Mahmudi, and Mohammad Hasan Naderi. On the maximal graph of a commutative ring. *Mathematics Interdisciplinary Research*, 2021.

[77] Masoumeh Soleimani and Mohamad Hasan Naderi. on power graph of some finite rings. *Mathematics Interdisciplinary Research*, 8(2):161–173, 2023.

[78] Sarah Spiekermann, Alessandro Acquisti, Rainer Böhme, and Kai-Lung Hui. The challenges of personal data markets and privacy. *Electronic markets*, 25(2):161–167, 2015.

[79] Sarah Spiekermann, Jens Grossklags, and Bettina Berendt. E-privacy in 2nd generation e-commerce: privacy preferences versus actual behavior. In *Proceedings of the 3rd ACM conference on Electronic Commerce*, pages 38–47, 2001.

[80] Stoyan R Stoyanov, Leanne Hides, David J Kavanagh, Oksana Zelenko, Dian Tjondronegoro, and Madhavan Mani. Mobile app rating scale: a new tool for assessing the quality of health mobile apps. *JMIR mHealth and uHealth*, 3(1):e27, 2015.

Age and dark pattern designs — 17[81] Fred Stutzman and Jacob Kramer-Duffield. Friends only: examining a privacy-enhancing behavior in facebook. In *Proceedings of the SIGCHI conference on human factors in computing systems*, pages 1553–1562, 2010.

[82] Mart Tacken, Fiorella Marcellini, Heidrun Mollenkopf, Isto Ruoppila, and Zsuzsa Szeman. Use and acceptance of new technology by older people. findings of the international mobilate survey:'enhancing mobility in later life'. *Gerontechnology*, 3(3):126–137, 2005.

[83] Monika Taddicken. The 'privacy paradox'in the social web: The impact of privacy concerns, individual characteristics, and the perceived social relevance on different forms of selfdisclosure. *Journal of Computer-Mediated Communication*, 19(2):248–273, 2014.

[84] Shalini Talwar, Amandeep Dhir, Puneet Kaur, Nida Zafar, and Melfi Alrasheedy. Why do people share fake news? associations between the dark side of social media use and fake news sharing behavior. *Journal of Retailing and Consumer Services*, 51:72–82, 2019.

[85] Lihong Tang, Wanlun Ma, Marthie Grobler, Weizhi Meng, Yu Wang, and Sheng Wen. Faces are protected as privacy: An automatic tagging framework against unpermitted photo sharing in social media. *IEEE Access*, 7:75556–75567, 2019.

[86] Jiang Tao and Hu Shuijing. The elderly and the big data how older adults deal with digital privacy. In *2016 International Conference on Intelligent Transportation, Big Data & Smart City (ICITBS)*, pages 285–288. IEEE, 2016.

[87] Zeynep Tufekci. Can you see me now? audience and disclosure regulation in online social network sites. *Bulletin of Science, Technology & Society*, 28(1):20–36, 2008.

[88] Amos Tversky and Daniel Kahneman. Rational choice and the framing of decisions. In *Multiple criteria decision making and risk analysis using microcomputers*, pages 81–126. Springer, 1989.

[89] Amos Tversky and Daniel Kahneman. Advances in prospect theory: Cumulative representation of uncertainty. *Journal of Risk and uncertainty*, 5(4):297–323, 1992.

[90] Evert Van den Broeck, Karolien Poels, and Michel Walrave. Older and wiser? facebook use, privacy concern, and privacy protection in the life stages of emerging, young, and middle adulthood. *Social Media+ Society*, 1(2):2056305115616149, 2015.

[91] Eka Dyar Wahyuni and Arif Djunaidy. Fake review detection from a product review using modified method of iterative computation framework. In *MATEC web of conferences*, volume 58, page 03003. EDP Sciences, 2016.

[92] Pei-Shan Wei and Hsi-Peng Lu. An examination of the celebrity endorsements and online customer reviews influence female consumers' shopping behavior. *Computers in Human Behavior*, 29(1):193–201, 2013.

[93] Marc G Weinberger and William R Dillon. The effects of unfavorable product rating information. *ACR North American Advances*, 1980.

[94] Pamela Wisniewski, AKM Islam, Heather Richter Lipford, and David C Wilson. Framing and measuring multidimensional interpersonal privacy preferences of social networking site users. *Communications of the Association for information systems*, 38(1):10, 2016.

[95] Pamela Wisniewski, Heng Xu, Heather Lipford, and Emmanuel Bello-Ogunu. F acebook apps and tagging: The trade-off between personal privacy and engaging with friends. *Journal of the Association for Information Science and Technology*, 66(9):1883–1896, 2015.

[96] Darrell A Worthy, Marissa A Gorlick, Jennifer L Pacheco, David M Schnyer, and W Todd Maddox. With age comes wisdom: Decision making in younger and older adults. *Psychological science*, 22(11):1375–1380, 2011.

[97] Heng Xu and Sumeet Gupta. The effects of privacy concerns and personal innovativeness on potential and experienced customers' adoption of location-based services. *Electronic Markets*, 19(2-3):137–149, 2009.

[98] Eva-Maria Zeissig, Chantal Lidynia, Luisa Vervier, Andera Gadeib, and Martina Ziefle. Online privacy perceptions of older adults. In *International Conference on Human Aspects of IT for the Aged Population*, pages 181–200. Springer, 2017.

# 9 Appendix

Table A1. A summary of the saturated path model including all of the significant findings. Please note that some of the results are different than what being reported in the hypotheses testing as this is the result of the saturated model rather than only hypothesized model (* $p < .05$, ** $p < .01$, *** $p < .001$)

| Variables | b (OR) | SE | p |
|---|---|---|---|
| **DV: Tagging Decision** | | | |
| **Age Group (OA vs. YA, H9)** | **0.160 (1.173) *** | **0.073** | **.028** |
| Privacy Concerns (H1) | 0.049 (1.050) | 0.087 | .573 |
| **Framing (positive vs. negative, H2)** | **0.359 (1.431) **** | **0.079** | **.0001** |
| **Default (opt-out vs. opt-in, H3)** | **0.184 (1.202) **** | **0.070** | **.009** |
| Justifications | $\chi^2(4) = 1.402$ | | .843 |
|     No vs. Any | -0.047 (0.954) | 0.080 | .559 |
|     Negative vs. Positive (H4) | -0.030 (0.970) | 0.078 | .706 |
|     Normative vs. Rationale | 0.058 (1.059) | 0.078 | .461 |
|     Justification type X Valence | 0.033 (1.033) | 0.081 | .686 |
| Age Group X Privacy Concerns | -0.136 (0.872) | 0.110 | .216 |
| **Age Group X Framing** | **0.185 (1.203) *** | **0.082** | **.029** |
| Age Group X Default | 0.189 (1.208) | 0.101 | .062 |
| Age Group X Justifications | $\chi^2(4) = 4.822$ | | .306 |
|     No vs. Any | -0.022 (0.978) | 0.104 | .833 |
|     Negative vs. Positive | -0.064 (0.938) | 0.086 | .454 |
|     Normative vs. Rationale | -0.158 (0.853) | 0.079 | .046 |
|     Justification type X Valence | -0.089 (0.914) | 0.087 | .306 |
| **DV: Privacy Concerns** | | | |
| Age Group (OA vs. YA, H8) | 0.093 | 0.059 | .117 |
| Framing (positive vs. negative, H5) | -0.013 | 0.078 | .867 |
| **Default (opt-out vs. opt-in, H6)** | **0.161** | **0.071** | **.024** |
| **Justifications** | $\chi^2(4) = 10.133$ | | **.038** |
|     No vs. Any | -0.038 | 0.072 | .600 |
|     **Negative vs. Positive (H7)** | **0.132 *** | **0.065** | **.042** |
|     Normative vs. Rationale | 0.040 | 0.067 | .551 |
|     **Justification type X Valence** | **-0.131 *** | **0.066** | **.046** |
| Age Group X Framing | 0.008 | 0.072 | .913 |
| **Age Group X Default** | **0.189 **** | **0.070** | **.007** |
| Age Group X Justifications | $\chi^2(4) = 3.707$ | | .447 |
|     No vs. Any | 0.070 | 0.057 | .217 |
|     Negative vs. Positive | 0.097 | 0.076 | .207 |
|     Normative vs. Rationale | 0.042 | 0.079 | .597 |
|     Justification type X Valence | 0.008 | 0.079 | .921 |